\documentclass[traditabstract]{aa}
\usepackage{graphicx}
\usepackage{wasysym}
\usepackage{txfonts}
\usepackage{natbib}
\usepackage{ulem}
\bibpunct{(}{)}{;}{a}{}{,} % to follow the A&A style
\usepackage{lscape}

\def\cone {\ifmmode{{\rm C}{\rm \small I}(^3\!P_1\!-^3\!P_0)}
     \else{C\ts {\scriptsize I}{\small$(^3\!P_1\!-^3\!\!\!P_0)$}}\fi}
\def\ctwo {\ifmmode{{\rm C}{\rm \small I}(^3\!P_2\!-^3\!P_1)}
     \else{C\ts {\scriptsize I}{\small$(^3\!P_2\!-^3\!\!\!P_1)$}}\fi}

\def\tex {\ifmmode{{T}_{\rm ex}}\else{$T_{\rm ex}$}\fi}
\def\tmb {\ifmmode{{T}_{\rm mb}}\else{$T_{\rm mb}$}\fi}
\def\ci     {\ifmmode{{\rm C}{\rm \small I}}\else{C\ts {\scriptsize I}}\fi}
\def\hi     {\ifmmode{{\rm H}{\rm \small I}}\else{H\ts {\scriptsize I}}\fi}
\def\hh     {\ifmmode{{\rm H}_2}\else{H$_2$}\fi}

\def\ts     {\thinspace}
\def\kms    {\ifmmode{{\rm \ts km\ts s}^{-1}}\else{\ts km\ts s$^{-1}$}\fi}
\def\msol   {\ifmmode{{\rm M}_{\odot}}\else{M$_{\odot}$}\fi}
\def\lsol   {\ifmmode{{\rm L}_{\odot}}\else{L$_{\odot}$}\fi}
\def\zsol   {\ifmmode{{\rm Z}_{\odot}}\else{Z$_{\odot}$}\fi}

\begin{document}

\title{On the steady state collisional evolution of debris disks around M~dwarfs}

\subtitle{}

\author{
Etienne Morey \inst{1}
\and
Jean-Fran\c cois Lestrade  \inst{2}
}
\institute{
Observatoire de Paris - LERMA, 61 Av. de l'Observatoire,  F-75014, Paris, France
\and
Observatoire de Paris - LERMA, CNRS, 61 Av. de l'Observatoire,  F-75014, Paris, France
}

\offprints{J-F Lestrade, \email{jean-francois.lestrade@obspm.fr}}

\date{Received ...... ; accepted ....}
\titlerunning{Steady state collisional evolution of  debris disks around M dwarfs}

\abstract{Debris disks have been found primarily around intermediate and solar mass stars (spectral types A-K), but rarely around low-mass M-type stars.
This scarcity of detections in M~star surveys can be confronted with the predictions of the steady state collisional evolution model. First, we determine 
the parameters of the disk population evolved with this model and fit to the distribution of the fractional dust luminosities measured in the surveys 
of A- and FGK-type stars observed by the infrared satellite {\it Spitzer}. Thus, in our approach, we stipulate that the initial disk mass distribution 
is bimodal and that only high-mass collisionally-dominated disks are detected. The best determined parameter is the diameter $D_c$ of the largest planetesimals 
in the collisional cascade of the model, which ranges between 2 and 60~km, consistently for disks around  A- and FGK-type stars. Second, we assume that 
the same disk population surrounds the M~dwarfs that have been the subjects of debris disk searches in the far-infrared with {\it Spitzer} and 
at submillimeter wavelengths with radiotelescopes. We find, in the framework of our study, that this disk population, which has been fit to the AFGK data, 
is still consistent with the observed lack of disks around M~dwarfs with {\it Spitzer}. }


\maketitle

%\citet{jon90}  :  Jones et al. (1990)
%\citep{jon90} 	:  (Jones et al. 1990)
%\citep[see][]{jon90} 	: (see Jones et al. 1990)
%\citep[see][chap.~2]{jon90}  : (see Jones et al. 1990, chap. 2)

%Multiple citations can be made as usual, by including more than one citation key in the 
% \cite command argument.
%\citet{jon90,jam91}    : Jones et al. (1990); James et al. (1991)
% \citep{jon90,jam91}   : (Jones et al., 1990; James et al. 1991)
%\citep{jon90,jon91} 	: (Jones et al. 1990, 1991)
%&é"'\citep{jon90a,jon90b} 	: (Jones et al. 1990a,b)

\section {Introduction} \label{intro}

A debris disk around a main sequence star is formed of planetesimals leftover 
from the planet formation process according to the core accretion theory. 
In our solar system, the Asteroid belt and Edgeworth-Kuiper Belt would be a debris disk 
to an hypothetical distant observer. Planetesimals are objects whose accretion was stymied by the formation and 
migration of giant planets in the system, or whose accretion simply occurred too slowly to grow larger.
The relationship between debris disks and planetary systems is debated and  should eventually place  
our solar system in context  \citep{Grea10}. 
The physical and observational  properties of debris disks  were defined by \citet{Lagr00}, and 
their studies were reviewed by \citet{Wyat08} and \citet{Kriv10}.

It is thought that observed dust associated with debris disks 
must be replenished through continual collisional grinding or sublimation
of planetesimals
resulting either from a steady state or a stochastic process, or both. 
About $10-30$\% of mature stars from stellar type A to K harbor cold dust disks
detected as  excess emission above photospheric levels in the far-infrared 
at $\lambda > 70~\mu$m. The observed decay 
of these excesses with star ages 
is seen as evidence of evolution under a steady state collisional process \citep{Domi03, Riek05, Bryd06, Wyat07a, Wyat07b,Carp09, Kain11}. 
The fact that debris disks are found around stars that have several orders of magnitude difference 
in stellar luminosities implies  that planetesimal formation (a critical step in planet formation) 
is a robust process that can take place under a wide range of conditions.
   
In addition to searches targeted at A- to K- type stars, 
several large surveys have been conducted to search for cold debris disks around M dwarfs 
in the mid-infared by \citet{Plav05, Plav09}, initially, and  \citet{Aven12}  with {\it WISE} (Wide Field Infrared Survey), 
as well as in the far-infrared 
with the satellites {\it Spitzer} and {\it Hershel} \citep{Gaut07, Matt10}, and in the submillimeter 
domain  with {\it JCMT} (James Clerk Maxwell Telescope)  and {\it IRAM30-meter} (Institut de Radioastronomie Millim\'etrique) \citep{Lest06, Lest09}. 
They have only yielded  a {\it bona fide} disk around the mature M3~dwarf GJ581 \citep{Lest12}, in addition to the
disk around the young M0~dwarf AU~Mic (12~Myr)  \citep{Liu04, Kala04}.  
 
The fact that debris disks are less frequently detected among M-dwarfs than around 
higher-mass stars seems surprising at first, since stars of all spectral types have similar detection 
frequency of protoplanetary disks in the earlier stages of their evolution \citep[e.g.,][]{Andr05}. 
It is also surprising in view of the
recent observations showing that small, low-mass planets are more  abundant  
among M-dwarfs than around stars of other types \citep{Howa11, Bonf13, Dres13, Kopp13}, and in light of 
the prevalence of debris disks around G-type stars hosting low-mass planetary systems 
(\citet{Wyat12},  Marshall et al., and Moro-Mart\'in et al. in prep).   
If the correlation between debris disks and low-mass planets for G stars 
applies to M~dwarfs, then debris disks should be relatively common around M~dwarfs, 
in contrast to recent observations.   

In reality, current observations may simply not be sensitive enough because  
 dust experiences significantly less heating around low-luminosity M~dwarfs 
than around more massive stars. However, M~dwarfs are usually closer
and may have larger total dust-emitting surfaces because small grains are not blown out of the
system by radiation pressure, unless a strong stellar wind is present.

In this paper, we show that under specific conditions (no stellar wind and 
high-mass collisionally-dominated disks),  the disk population around M~dwarfs is not
different from the one around other types of stars, despite the lack of detected debris disks in surveys. 
In \S \ref{pop}, we describe the steady state collisional evolution model developed in \citet{Wyat07a} 
and our adaptation for its use. In \S \ref{Deter}, we fit this model 
to the data of the {\it Spitzer} surveys of A~stars and  FGK~stars with a novel fitting 
procedure, assuming a bimodal distribution for the masses of the
debris disks. Interestingly, 
the best-fits of the two samples are shown to be consistent.  
Finally, in \S \ref{Mdwarfs}, we use this model and its assumptions
to show that the observed lack of debris disks around M~dwarfs found in the survey of \citet{Gaut07} in the far-infared domain
is consistent with the observations reported for A and FGK~stars.

\section {Model for the population of debris disks} \label{pop}

\subsection{Analytic steady state collisional evolution model} \label{mod}

We assume that stars with detected debris disks are surrounded by a 
belt of planetesimals that is undergoing collisional cascades producing the dust 
observed. The long-term evolution of such a belt in steady state collisional equilibrium 
was originally developed analytically in \citet{Domi03} and recast with slightly 
modified assumptions in \citet{Wyat07a}. This model shows that debris disks decay 
at a constant rate proportional to $t^{-1}$. An extension of this analytic model, which also 
includes the evolution of the star into the subgiant phase, is presented in \citet{Bons10}, and yields a similar decay.
Numerical models of single systems have shown that this decay is better described as
a quasi-steady state with rates varying with time \citep{Theb03, Lohn08, Theb07, Gasp12a}.    

In this study, we use the analytic model of \citet{Wyat07a} for its simplicity and
overall accuracy in modelling collision-dominated disks that have been observed in
surveys so far.  In this model, the 
collisional equilibrium is described with a single power law for the size distribution defined by 
$n(D)=K D^{2-3q}$ between $D$ and $D+dD$, and with $q=11/6$ \citep{Dohn69}. Such a collisional cascade implies infinity at both
ends. But in debris disks, the cascade is truncated 
at the top end by the largest planetesimal of diameter $D_c$, and at the 
lower end by the blowout size if the stellar radiation pressure can blow out 
the smallest particles. 
In this model, the size distribution is independent of radial distance in the disk. The coefficient $K$
can be related straightforwardly to the total mass or the total surface of the material. 
We recall that for index $q=11/6$, most of the mass of the belt is in the largest planetesimals, 
while most of the emitting cross-sectional area is in the smallest particles.

In a collisional cascade, material within a given size range $D$ to $D+dD$ is
replaced by fragments from the destruction of larger objects at the same
rate that it is destroyed in collisions with other members of the cascade. The 
long timescale evolution is thus determined by the removal of mass from the
top end of the cascade, and, hence, by the collisional lifetime $t_c(0)$ of
the largest planetesimals of size $D_c$ at the initial epoch. 
In \citet{Wyat07a}, planetesimals of size $D_c$
are destroyed by impacting planetesimals down to size $X_c D_c$  ($X_c < 1$)  making
collisional lifetimes shorter than in \citet{Domi03} where the large
planetesimals feeding the cascade are less realisticaly treated, as a population separate 
from the cascade. The ratio $X_c = D_{cc}/D_c$, where $D_{cc}$ is the 
smallest planetesimal that has enough energy to catastrophically 
destroy a planetesimal of size $D_c$,  can be expressed from the dispersal threshold 
$Q_D^*$, {\it i.e.}, the specific incident energy required to catastrophically 
destroy a particle.  Large bodies 
are eventually ground to small particles generally expelled from the system 
by  radiation or wind pressure.

However, for low luminosity M~dwarfs, the smallest grains cannot be expelled by radiation 
pressure, but by
stellar wind if significantly above the solar value  \citep{Plav05}. This might be the case
for very young M dwarfs, such as 
AU Mic, and possibly for fully convective stars later than the substellar spectral type M3
\citep{Hawl00, Warg01}. 
The M dwarfs of unknown ages in the Spitzer sample of \citet{Gaut07} are likely older
than 1~Gyr,  hence we have assumed  no stellar wind
to simplify their study and to test other parameters of the model,
especially planetesimal size. 
We have set the smallest particle diameter $D_{min}$   for M dwarfs
to $0.1~{\mu}$m  in our model  because the absorption efficiency $Q_{abs}$ drops 
significantly below unity for smaller grains \citep{Laor93} making their emission negligible.   

In the model,
the total mass $M_{tot}(t)$ of the planetesimals in the belt decays as:

\begin{equation}
  M_{tot}(t) = {{M_{tot}(0)} \over {[1 + t/t_c(0)]}} \label{eq:Mt}
\end{equation}

\noindent where $M_{tot}(0)$ is the initial total mass  
and $t_c(0)$ is the collisional time scale at that initial epoch.
This solution is valid as long as the mass is the only time-variable 
property of the disk. 
For the collisional lifetime $t_c(0)$, we used Eq.~9 of \citet{Wyat07b} summarized here by:

\begin{equation}
 t_c(0) \propto {{\rho R^{3.5} \times \Delta R/R \times D_c} \over {M_*^{0.5} M_{tot}(0)}}~f(X_c,e/I,q) \label{eq:tc0}
\end{equation}

\noindent  and their expression for the function $f$. The factor $X_c$ has been derived in \citet{Wyat02} :

\begin{equation}
X_c \propto \Big({{Q_D^*(D_c) \times R \times M_*^{-1}} \over { 1.25 e^2 + I^2}}\Big)^{1/3} \label{eq:Xc}
\end{equation}

\noindent These expressions depend on the conditions in the disk 
(total mass $M_{tot}(0)$ unless $ t >> t_c(0)$, mean radius $R$ and width $\Delta R$ of the disk, 
diameter $D_c$ of the largest planetesimals) and on the parameters controlling  
its subsequent collisional evolution
(eccentricity and inclination ($e, I$)  of the perturbed planetesimals, 
material density $\rho$, and the  star mass $M_*$).

The specific energy threshold $Q_D^*$ for catastrophic disruption depends on the largest
planetesimal $D_c$, and we use the model of \citet{Benz99}:

\begin{equation}
Q_D^*(D_c)= Q_0 (D_c/2)^a + B \rho  (D_c/2)^b \label{eq:thres}
\end{equation}

\noindent where $Q_0 \sim 1.6~10^7 - 9~10^7$~erg g$^{-1}$, $B \sim 0.3 - 2.1$~erg~cm$^{-3}$~g$^{-2}$, 
$a \sim -0.4$, and $b \sim 1.20$, as given in their Table~III for basalt and ice, and 
impact velocities 0.5, 3 and 5~km/s. 
The assumptions that $Q_D^*$ and $D_c$ are independent of stellar age 
and are the same for all disks are limitations of the model.    

We synthesize a disk population by adopting distributions for 
the initial disk masses $M_{tot}(0)$ and the mean disk radii. The distribution for $M_{tot}(0)$
is based on the submillimeter study of protoplanetary disks of \citet{Andr05} who derived
a lognormal distribution of their dust masses  centered on $M_{mid}=15~M_{\oplus}$ 
(standard mass opacity of 0.1~cm$^2$~g$^{-1}$ at 1000~GHz and mass ratio of 100:1 between gas and dust)
and with a $1\sigma$ width of 0.71~dex. 
As described later, we fit for  $M_{mid}$ but retained this  $1\sigma$ width 
in our adjustment of the debris disk data.
 The distribution for the mean disk radii $R$ is taken as
the power-law $n(R) \propto R^{\gamma}$ between $R$ and $R+dR$ and is bound by the parameters 
$R_1$ and $R_2$. We fix $\gamma$ to $-0.5$ 
based on the distributions found for the characteristic radii of the protoplanetary disks
 measured by the Submillimeter Array (SMA) in \citet{Andr11} and on the distribution of the mean radii of the nine resolved debris disks 
around A-stars of the Herschel program DEBRIS (Disc Emission via a Bias-free Reconnaissane in the Infrared/Submillimetre) \citep{Boot13}.  

The total mass $M_{tot}(t)$ of colliding planetesimals at the age of the star can be converted into 
the total cross-sectional area $A(t)$ dominated by the dust with a 
mass density $\rho$ and the standard size distribution  
$n(D) \propto D^{-3.5}$ between $D$ and $D+dD$ for spherical particles of diameter $D$
from the largest value $D_c$ to the smallest $D_{min}$ taken as  0.1~$\mu$m (M dwarfs) 
or as the blowout size if larger than  0.1~$\mu$m ({\it i.e.}, most other stellar types).
The total cross-sectional area $A(t)$ at the age of the star is:

\begin{equation} 
A(t) = {{3 \times M_{tot}(t)}  \over {2 \rho \sqrt{D_{min} D_c}}}.
\label{eq:A}
\end{equation}

Finally, the fractional dust luminosity:

\begin{equation}
  { f_d}(t)= {L_{dust} \over L_*}  \label{eq:fd1}  
\end{equation}

\noindent is calculated  taking into account that the dust grains are spatially distributed according to a radial profile 
taken as the power-law $\Sigma_p r^{\alpha}$ between $r$ and $r+dr$ 
of the radial extent $r$ from  
the inner radius $r_{in}$ to the outer radius $r_{out}$ of the disk. 
The term $\Sigma_p r^{\alpha}$ is the emitting cross-sectional 
area of the grains per unit area of the disk surface. In assuming that 
these grains absorb stellar light with 100\% efficiency $(Q_{abs}=1)$, their total emission is
proportional to $A(t) = \int_{r_{int}}^{r_{out}}  2 \pi r dr \Sigma_p r^{\alpha}$. Hence, 
if $\alpha \ne 0$ and using Eq~(3) of \citet{Lest12} for $\Sigma_p$, 
the fractional dust luminosity is:

\begin{eqnarray}
{f_d(t)} & = & \int_{r_{in}}^{r_{out}} {{2 \pi r dr \Sigma_p r^{\alpha}} \over {4 \pi r^2}}  = \nonumber \\  
        &  &  {{1}\over {2 \pi}} {{A(t) \times (\alpha +2)} \over {\alpha}}  \times 
\Big[{{r_{out}^{\alpha}-r_{in}^{\alpha}} \over {r_{out}^{\alpha+2}-r_{in}^{\alpha+2}}}\Big].
\label{eq:fd2}
\end{eqnarray}

\noindent  This expression of the fractional dust luminosity accounts  
for the radial extent of the disk, while it is usually estimated at its mean radius. 
The corresponding change is about 50\% for a disk width of $\Delta R=R/2$ and $\alpha=-1.5$, as adopted in this study. With these notations,
the inner and outer radii in Eq~(\ref{eq:fd2}) above are $r_{in}=R-\Delta R/2$ 
and  $r_{out}=R+\Delta R/2$, respectively.
As mentioned above, we adopt a power-law distribution for the mean disk radii $R$ of  the modeled 
disk population extending 
between the lower and upper boundaries  $R1$ and $R2$. 
These boundaries must not be confused with $r_{in}$ and   $r_{out}$ in Eq~(\ref{eq:fd2}).

\subsection{Fitting procedure} \label{fit}

\null

The searches for debris disks around  A stars \citep{Su06} and
FGK stars \citep{Tril08} at 24 and 70~$\mu$m with {\it Spitzer} are based on large samples of stars
that are statistically not well defined   because they are assembled from several programs
based on  various selection criteria  and  different integration times.
Consequently, these surveys are neither volume-limited, nor flux-limited, and not
representative of the  stellar population  in the solar neighborhood. For instance, the sample
in \citet{Su06} made of stars with spectral types ranging from B6 to A7 is highly biased toward
A0 stars 
that constitute 27\% of the selection while the  A6 and A7 stars amount to less than 5\%, unlike 
the actual stellar density in the solar neighborhood. The FGK star sample in  \citet{Tril08} is 
biased toward F5-type stars representing 16\% of this sample, 
while all K0 to K4 stars account for only 10\%, 
unlike the stellar population in the solar neighborghood.
In these conditions, it is debatable whether or not the detection frequency estimated from these 
samples are statistically 
representative of the population of cold outer debris disks as assumed
in previous evolution studies. 

In this work, we  turn to the possibility that 
the distribution of their total masses is bimodal. High-mass cold disks 
are dense enough to  be collisionally dominated and thus dusty enough to
be detectable with {\it Spitzer} in the far-infrared and radiotelescopes
at longer wavelengths. Instead, low mass 
cold  disks are Poynting-Robertson-dominated and below detection level at these wavelengths, and
they are possibly feeding exozodical regions detectable at much
shorter wavelengths \citep{Wyat05}. There are several reasons for this possible bimodal 
distribution.
A fraction of outer debris disks can be stripped of their planetesimals during 
close stellar flybys
while in the open cluster of their birth for the first 100~Myr of their lifetime 
if the stellar density is high \citep{Lest11}. Also,
planet-planet gravitational scattering can trigger dynamical instabilities 
that can eject planets and  drastically remove planetesimals  any time 
during the lifetime of the system. 
This scattering model is supported by several features of the
observed giant exoplanets, notably their broad eccentricity
distribution \citep{Chat08, Juri08, Raym10} and 
their clustering near the Hill stability
limit \citep{Raym09}.
Finally, giant planets, in crossing their mean motion resonances during migration,
can destabilize the outer disk, triggering the Late Heavy Bombardment and
ejecting planetesimals as in the Nice model for the solar system  \citep{Gome05, Morb05, Tsig05}.

Hence, in this work, for each survey,
we have only retained the stars with detected debris disks and made the hypothesis
that this subsample is statistically representative of the population of 
high-mass cold disks. The nondetections are stars with
low-mass disks that cannot be observed with our current capabilities in the 
far-infrared and
submillimeter. Our statistical results hold within this framework.  

For such a subsample, stellar distances, ages, and spectral 
types are known and only the parameters of the modeled debris disk population are unknown. 
In our study, this disk population is synthesized with
the steady state collision model described in the previous section. 
This model is evolved using the known or adopted stellar ages 
and is fit to the distribution of observed fractional dust luminosities of the detected disks.  
The advantage of using the fractional dust luminosity rather than the photospheric excesses 
is that this quantity is distance-independent and  combines both the 24 and
70~$\mu$m photometric measurements of the survey.   

In our study, the stellar ages and spectral types, {\it i.e.}, the masses
and luminosities of the stars,  are taken from  \citet{Su06}
and \citet{Tril08}. The initial total mass of planetesimals
$M_{tot}(0)$ and the mean disk radii $R$ for
the detected disks of the subsample  were randomly drawn from their distributions.  
As justified in \S \ref{mod}, we adopted a log-normal distribution for $M_{tot}(0)$ 
centered on  the median mass 
$M_{mid}$ and having a $1\sigma$ width $\sigma_{M_{tot}}$ of 0.71~dex, 
and adopted the power-law distribution $R^{\gamma}$ from $R1$ to $R2$, 
with $\gamma=-0.5$, for the mean disk radius $R$. 
In total, there are four independent parameters in our fit : $M_{mid}$, 
$R1$, $R2$, and the diameter $D_c$ of the largest planetesimals. 
We have set fixed $e=I=0.05$, $\rho=1000$~kg~m$^{-3}$ (ice) or 
$\rho=2700$~kg~m$^{-3}$ (basalt), $\Delta R=R/2$, and $q=11/6$ 
for all disks  as in past studies of \citet{Wyat07b} and \citet{Kain11}. 
We have considered  the effects of changing these parameters in \S \ref{Deter}.

The best-fit values of the four parameters
$M_{mid}$, $D_c$, $R1$, and $R2$ of the model were searched numerically over a large grid  using 
the Kolmogorov-Smirnov test (K-S test) to match the observed  
and modeled distributions of fractional dust luminosities. At each point of this 
4D-parameter space, 100 simulations were carried out so that  $M_{tot}(0)$  and $R$ 
take many randomly chosen values from their adopted distributions 
as in the Monte Carlo method. The best-fit values were 
those yielding the highest K-S test probability averaged over the 100 simulations.

\begin{figure}[h!]
%\resizebox{8.5cm}{!}{\includegraphics[angle=-90] {fit_Su06.ps}}
\resizebox{8.5cm}{!}{\includegraphics[angle=-90] {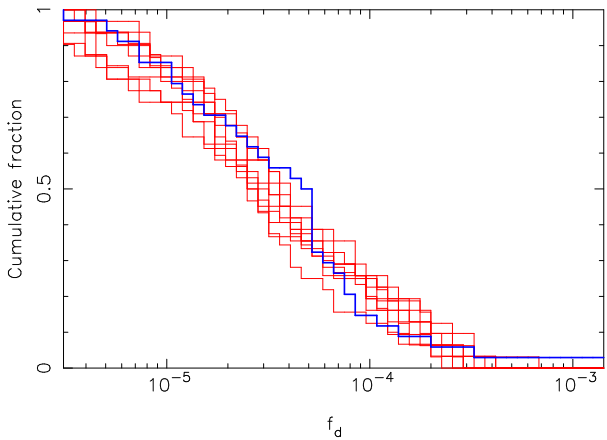}}
\caption{Cumulative distributions of the fractional dust luminosities; the blue line corresponds to 
the  observations 
and the red lines are eight independent simulations computed with the same parameters  
$M_{mid}$, $R1$, $R2$, and $D_c$, but with different random numbers to simulate 
the disk population around 
the A stars of the \citet{Su06} survey. The statistical Kolmogorov-Smirnov test is used to match 
the observed and simulated distributions so they can be considered drawn from the same parent 
distribution. The maximum K-S probability is 75\% in 
this example, averaged over 100 simulations.  For clarity, only eight simulations are shown
on this figure but 100 were actually computed at each point of the 4-D parameter space ($M_{mid}$, $R1$, $R2$, $D_c$) for our numerical search.} 
\label{fig:cumul}
\end{figure}

The K-S test measures the largest 
absolute difference between the two distributions under consideration and
estimates the probability of finding this difference larger than the observed value in the case of the null
hypothesis (data drawn from the same distribution).  
Thus, this probability is an estimate of the significance  level of the observed difference
as a disproof of the null hypothesis; {\it i.e.}, a small probability implies that the two distributions
do not come from the same parent distribution. We have used the surbroutines in \citet{Pres92} which  are
reliable when the number of data points is larger than 20 and therefore appropriate for the subsamples
of debris disks in our study. Figure \ref{fig:cumul} is an illustration of our fitting procedure.

The values for $M_{tot}(0)$ and $R$ were drawn from the log-normal and power-law distributions, respectively, using 
the reciprocal of a random variable \citep[][]{Devr86}:

\begin{equation}
  x= \int_0^y f(t) dt      
\end{equation}

\noindent where $x$ is a random variable drawn uniformly between 0 and 1 and 
$y$ is an implicit variable that has  the density $f(t)$ that can be  
 set analytically or as a binned distribution   
\footnote{The variable $y$ becomes explicit when the density $f(t)$ is  
the power-law  $r^{\gamma}$ between $a$ and $b$ : 
$y = [(b^{\gamma +1} - a^{\gamma +1}) x + a^{\gamma+1}]^{1 \over {\gamma +1}}$}.

\subsection{Tests of the fitting procedure} \label{test}

\begin{figure}[h!]
%\resizebox{8.5cm}{!}{\includegraphics[angle=-90] {test1_b.ps}}
\resizebox{8.5cm}{!}{\includegraphics[angle=-90] {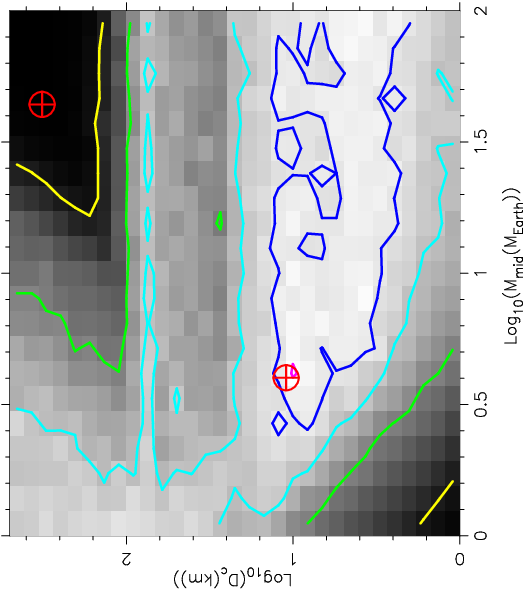}}

\null
\null

%\resizebox{8.5cm}{!}{\includegraphics[angle=-90] {test3_b.ps}}
\resizebox{8.5cm}{!}{\includegraphics[angle=-90] {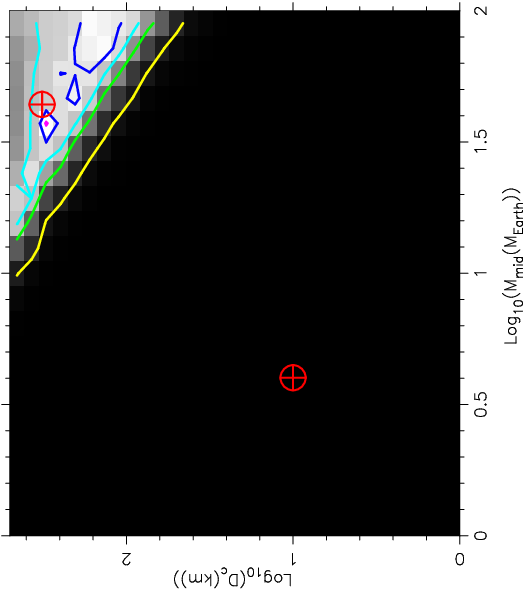}}
\caption{Tests of the fitting procedure with simulated data. 
{\it Top}~: parameter search for data simulated with $M_{mid}=4$~M$_{\oplus}$, $D_c=10$~km, 
$R1=6.5$~AU, $R2=35$~AU. {\it Bottom}~: parameter search for data simulated 
with $M_{mid}=40$~M$_{\oplus}$, 
$D_c=350$~km, and $R1$ and $R2$ as just mentioned. Levels are 99\% (red), 90\% (dark blue), 
75\%(light blue), 50\% (green) and 10\% (yellow). The two best-fit regions 
delineated by the 90\% contour  are  clearly exclusive when comparing the two
plots. Their peaks are close to 
the {\it a priori} values of $M_{mid}$ and $D_c$. The two crossed circle symbols on each plot 
indicate the positions of the {\it a priori} values used for 
$M_{mid}$ and $D_c$ to simulate the two sets of data.} 
\label{fig:test}
\end{figure}

We have simulated fractional dust luminosities computed with the collisional equilibrium evolution 
model of \citet{Wyat07a} described above, including
the relationship bewteen $Q_D^*$ and $D_c$ of \citet{Benz99}. In addition, a uniformly distributed 
random noise of $\pm 20 \%$ was added to each modeled fractional dust luminosity.
A first set of data was simulated
with the {\it a priori} values $M_{mid}=4$~M$_{\oplus}$ and $D_c=10$~km, and a second with
$M_{mid}=40$~M$_{\oplus}$ and $D_c=350$~km, to verify that these two disk populations,
which are clearly distinct, can be distinguished with our fitting procedure. 
We kept the other two fitted parameters
set to the same {\it a prioiri} values, $R1=6.5$~AU and $R2=35$~AU, for this verification.
The distributions of the stellar masses and ages were those of the sample of \citet{Su06}. 
In addition to the fitted parameters, the disk population was controlled
with the fixed parameters : 
$e=I=0.05$, $\Delta R=R/2$, $\gamma=-0.5$, $\sigma_{M_{tot}}=0.71$~dex 
and ice for the planetesimal material. The results of the fits 
are in Fig~\ref{fig:test} where the contours of the K-S test maximal probability indicate that
the best-fit areas delineated with the 90\% contours, although large, do not ovelap 
for the two data sets of our tests. It is noticeable that
the peaks of these regions are close to the {\it a priori} values of the parameters $M_{mid}$
and $D_c$ of the two simulated data sets. Hence, the two disk populations can be 
distinguished.
The impact of modifying the fixed parameters are studied with the real data in \S~\ref{Deter}.

\section {Determination of the parameters of the modeled debris disk population} \label{Deter}

\subsection {Parameter search for the A-star survey of \citet{Su06}}   \label{Para_A}

A total of 160 A stars with ages between 5 and 850 Myr and distances between 2.6 and 384 pc
was surveyed  at 24 and 70~$\mu$m with {\it Spitzer} \citep{Su06}. 
We have fit the model
to the distribution of the fractional dust luminosities of the sole 34 stars with 
detected disks at both 24~$\mu$m and 70~$\mu$m, and older than 9~Myr.  
We adopted the fractional dust luminosities $f_d$ of Table~3 (Group~I) of \citet{Su06} that are based 
on the observed [24]-[70] 
color temperature and on the assumption of blackbody emission for the dust.  
Since fractional dust luminosity
is only $\sim$ 8\% overestimated between a blackbody  and modified blackbody emission for A stars, 
this is irrelevent for our purpose. 
We did not retain the additional eight stars with only 70~$\mu$m excesses  (Group~II in \citet{Su06})
which yield solely  upper limits for $f_d$.

\begin{figure}[h!]
%\resizebox{7.75cm}{!}{\includegraphics[angle=-90] {b_A_ice_basic_sM_0_71_500km_1UA_pap_a.ps}}
\resizebox{7.75cm}{!}{\includegraphics[angle=-90] {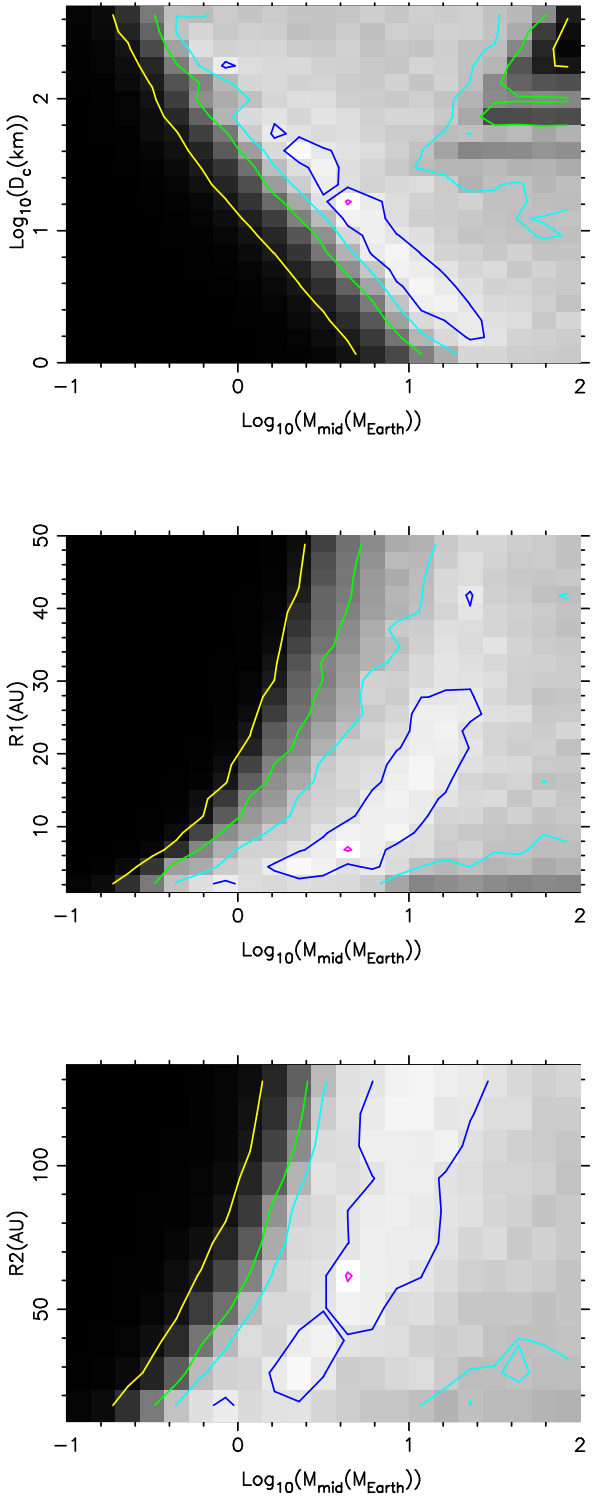}}
\caption{Best-fit regions for the modeled disk population around the A-type stars of the \citet{Su06} survey. 
Plots of the significance levels of the Kolmogorov-Smirnov test for the adjusted parameters
$M_{mid}$,  $D_c$, $R1$, and $R2$ of the model. 
The adopted values are 
$e=I=0.05$, $\Delta R=R/2$, $\gamma=-0.5$, $\sigma_{M_{tot}}=0.71$~dex 
and ice for the planetesimal material.
Levels are 99\% (red), 90\% (dark blue), 
75\%(light blue), 50\% (green), and 10\% (yellow) of the peak K-S test probability (80.7\%).
The $M_{mid}$ and $D_c$
are on a log scale, while R1 and R2 are on linear scale.} 
\label{fig:Su06}
\end{figure}

The ranges of the model parameters tested were : $M_{mid}$ of 0.1 to 100 $M_{\oplus}$ ; 
$D_c$ of 1 to 500~km ; $R1$ of 1  to 50~AU ; and  $R2$ of $R1$ + 10~AU to 150~AU.   
The grid was set up with 20 increments for each parameter. The fixed parameters are
the mean planetesimal eccentricity $e$=0.05, the disk width $\Delta R=R/2$, the exponent  
of the mean disk radius power-law distribution $\gamma= -0.5$, 
the $1\sigma$ width of the log-normal distribution of $M_{tot}(0)$ taken to be 0.71~dex, 
and ice for the planetesimal material ($Q_D^*$). 
We carried out a numerical search in these conditions and plots
of the significance levels based on the K-S test probability 
for each  pair of parameters are displayed in Fig~\ref{fig:Su06}. 
These plots indicate that probabilities are equally high  
over a broad region where the full ranges of the other two parameters are explored.  
They are not the 2-D projections of 
the 4-D parameter space for the maximum K-S test probability.
Such projections  provide ``peaked'' plots that are misleading because they 
falsely convey the impression that the solution is unique.

\begin{figure}[h!]
\resizebox{7.75cm}{!}{\includegraphics[angle=-90] {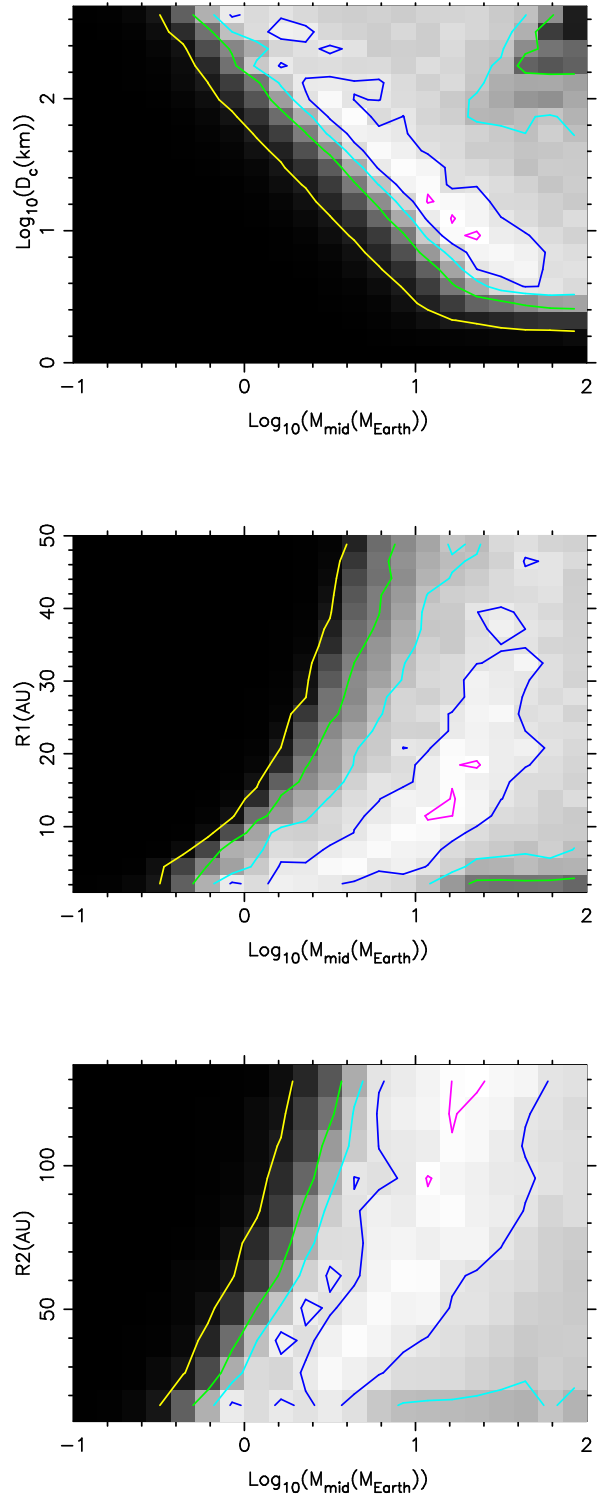}}
%\resizebox{7.75cm}{!}{\includegraphics[angle=-90] {b_A_ice__sM_0_71_500km_1UA_eta_10_pap_a.ps}}
\caption{Best-fit regions for the sample of A-type stars with a model assuming narrow disks ;
width is $\Delta R=R/10$ while it is $\Delta R=R/2$ in Fig \ref{fig:Su06}.  The peak K-S test probability 
is 77.6\% in this new fit. See the legend of Fig \ref{fig:Su06}  for more details. Compare the 90\% lower and
upper limits for M$_{mid}$ and $D_c$ relative to  Fig \ref{fig:Su06}.} 
\label{fig:Su06_bis}
\end{figure}

The most probable ranges for the parameters can be determined using the
90\% significance level in Fig~\ref{fig:Su06}. 
For the median mass $M_{mid}$, the lower and upper limits are  
2~M$_{\oplus}$  and  20~M$_{\oplus}$ ;  for $D_c$, they
are 2~km and 60~km ; for $R1$, they are 2~AU and 20~AU. For $R2$, the range is from 
30~AU to beyond $>150$~AU in Figs~\ref{fig:Su06},
\ref{fig:Su06_bis}.
Hence, the upper boundary $R2$ of the distribution of the mean disk 
radii is uncertain and can accommodate large disks, 
as a few have been observed ; for instance, the disk
around HD~207129, which has also been modeled with a collisional cascade
by \citet{Loeh12}.
Based on the values given above for $D_c$, the corresponding limits for $Q_D^*$ are between 62~J~kg$^{-1}$ 
and  895~J~kg$^{-1}$ for ice and are  between 92~J~kg$^{-1}$ and 2277~J~kg$^{-1}$ 
for basalt using the model of \citet{Benz99}.

The parameters $R1$ and $R2$ are blackbody radii because the fractional dust luminosities used were
estimated with this assumption for the grains.
Hence, the true disk radii should be about twice these values  in applying the factor $\Gamma=2$ (ratio of the
resolved radius to the blackbody radius) found for the resolved A star disks in the {\it Herschel}
DEBRIS program \citep{Boot13}.

The best-fit values in  the determination of \citet{Wyat07b}, based on a fit to the 
disk detection frequencies binned in ages and excess strengths, are  
$M_{mid} = 10$~M$_{\oplus}$, $D_c=60$~km, $Q_D^*$=300~J/kg (independent parameter in  their study), 
$R1=3$~AU, and  $R2$=120~AU. 
Additionally, they fit $\gamma=-0.8\pm 0.6$ consistent with our adopted value
of $-0.5$, and they assumed $e=I=0.05$ and $1\sigma$ width of 1.14~dex for the log-normal 
distribution of $M_{tot}(0)$. Their solution is within the 75\% significance level region 
of our fit in  Fig~\ref{fig:Su06}. 

We have changed the fixed parameters of the model to test whether or not
the best-fit regions are significantly modified. We changed the dynamical excitation of the disks 
$e=I$ (0.05/2, $0.05 \times 2$, $0.05 \times 4$),  we tested narrower disks with $\Delta R=R/10$, 
as well as shallower and steeper distributions for the mean disk radii $R$ with $\gamma$ ($0.0, -1.0$), 
we modified the $1\sigma$ width of the log-normal 
distribution of $M_{tot}(0)$ ($\sigma_{M_{tot}}=$ 0.25, 0.50, 1.14), and we tested the material 
of planetesimals (ice, basalt).  The main
structure and extent of the significance levels in the probability plots of Fig~\ref{fig:Su06_bis} are 
very similar to those of our standard  model ($e=0.05$, $\Delta R=R/2$, $\gamma=-0.5$,
$\sigma_{M_{tot}}=0.71$~dex, ice) in Fig~\ref{fig:Su06}. The most noticeable 
difference occured when changing from wide ($\Delta R=R/2$) to narrow ($\Delta R=R/10$) disks. 
Then, as shown in Fig~\ref{fig:Su06_bis}, the best-fit region is shifted upward, 
so that the lower and upper limits become $\sim$ 1~M$_{\oplus}$ and   
60~M$_{\oplus}$ for $M_{mid}$ and  6~km and 100~km for $D_c$, respectively. This can be explained
with Eq~(\ref{eq:tc0}) where it is clear that the collisional timescale  $t_c(0)$
controlling disk evolution is  proportional to $\Delta R/R \times D_c$, 
so narrower disks can be outweighed by increasing $D_c$.
     
Finally, we note that our optimum solution yields a slightly better K-S test probability (80.7\%) 
than when we changed the fixed parameters to other values. 
For instance, changing $\sigma_{M_{tot}}$ to 0.25, 0.50, and 1.14 dex
makes this probability becomes 77.7\%, 78.1\%, and 74.1\%,  respectively. 
 
\subsection {Parameter search for the solar type star survey of \citet{Tril08} } \label{Para_FGK}

Nearly 200 solar type stars (FGK) with ages between 290~Myr and 11.75 Gyr and distances 
between 4.7 and 148 pc were surveyed at 24 and 70~$\mu$m with {\it Spitzer} \citep{Tril08}.  
The sample was assembled from different programs.
For the same reason as for the A~star sample, we  fit the model
to the distribution of the fractional dust luminosities of  the sole 28 stars with detected disks.
The fractional dust luminosities of this subsample are taken from Tables~5 and 6 of \citet{Tril08}.
They are estimated from  the  [24]-[70] color temperature
when excesses were measured  at both bands (six disks). However, most excesses were detected at only 70~$\mu$m
and so these authors set the 24~$\mu$m ``excess'' to be equal to three times
the uncertainty at 24~$\mu$m and found the blackbody emission that best fit this excess and
the 70~$\mu$m excess. Such a temperature is the maximum temperature 
for the excess yielding the maximum fractional dust luminosities that are consistent with the data.

\begin{figure}[h!]
\resizebox{7.75cm}{!}{\includegraphics[angle=-90] {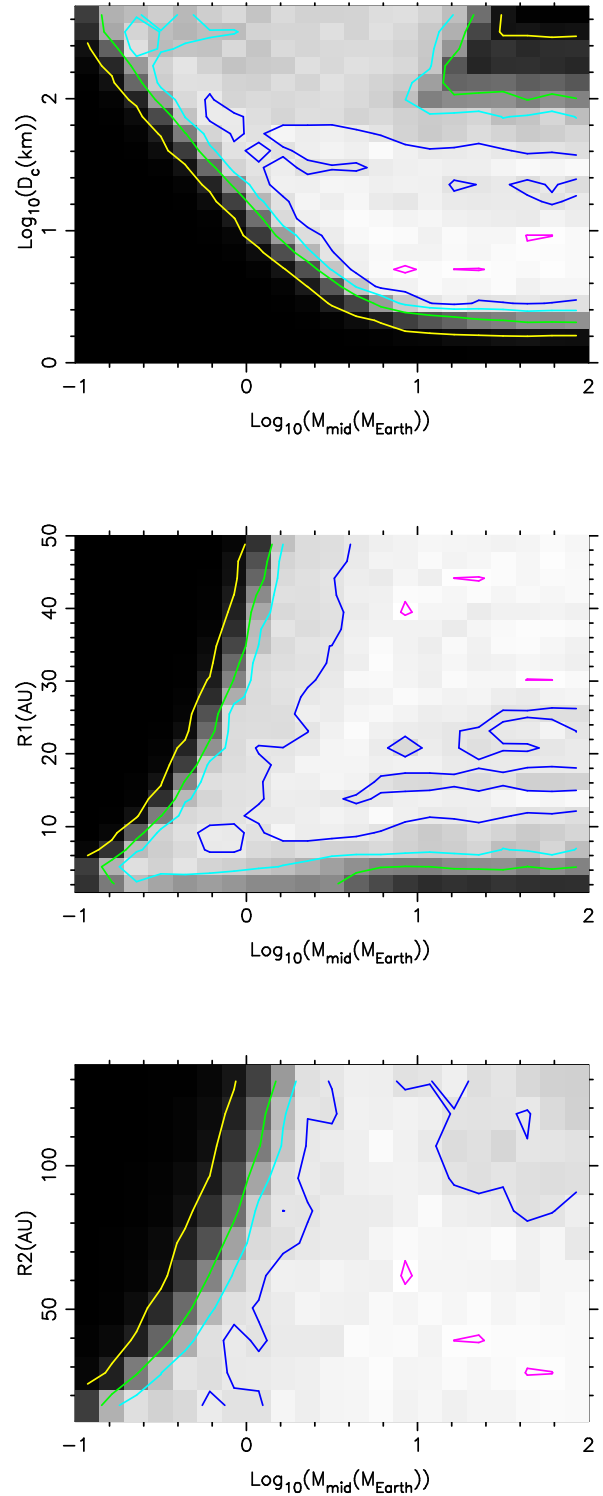}}
%\resizebox{7.75cm}{!}{\includegraphics[angle=-90] {b_F_ice_basic_sM_0_71_500km_1UA_pap_a.ps}}
\caption{Best-fit for the FGK-type stars of the \citet{Tril08} survey. 
Plots of the significance levels of the K-S test for the adjusted parameters $M_{mid}$,  $D_c$, $R1$, and $R2$ of the model. 
The adopted values are 
$e=I=0.05$, $\Delta R=R/2$, $\gamma=-0.5$, $\sigma_{M_{tot}}=0.71$~dex 
and ice for the planetesimal material. Levels are 99\% (red), 90\% (dark blue), 
75\%(light blue), 50\% (green), and 10\% (yellow) of the peak K-S test probability (87.1\%).
The $M_{mid}$ and $D_c$ are on a log scale, while R1 and R2 are on linear scale.} 
\label{fig:Tril08}
\end{figure}

The ranges of the parameters tested and the values adopted for the fixed parameters 
are the same as for the A~stars in \S \ref{Para_A}. Similarly, Fig~\ref{fig:Tril08} 
shows also regions of maximum K-S test probabilities marginalized 
(not 2-D projections of the 4-D parameter space). 
Regions of highest K-S test probability shown by the 90\% significance level in this figure are wider than
for the A~stars.

The FGK subsample is smaller and the fractional dust luminosities are mostly the maximum values from excesses detected 
at the single wavelength $\lambda = 70~\mu$m. This is a limitation
of the analysis and is likely responsible for a somewhat degraded fit of this subsample compared 
to the A stars in the previous section.
The 90\% significance level in Fig~\ref{fig:Tril08} yields 
only a lower limit of 0.8~M$_{\oplus}$ for $M_{mid}$ and  lower
and upper limits of 2~km and 60~km for $D_c$, assuming ice for the material composition. 
The parameters $R1$ and $R2$ are not well determined.

The best-fit values in the determination of \citet{Kain11}, based on a fit to the disk
detection frequencies binned in ages and excess strengths  of FGK stars, 
are  $M_{mid} = 4$~M$_{\oplus}$, $D_c=450$~km, $Q_D^*$=3700~J/kg 
(independent parameter in  their study), $R1=1$~AU, and  $R2$=160~AU. 
Additionally, they fit $\gamma=-0.60\pm 0.35$ and this is consistent with our adopted value
of $-0.5$, and they assumed $e=I=0.05$ and a $1\sigma$ width of 0.8~dex for the log-normal 
distribution of $M_{tot}(0)$. This solution is within the 50\% significance level region 
in our Fig~\ref{fig:Tril08}. 

In a similar fashion as for the  A~stars  discussed in the previous section, we changed the fixed parameters 
of the model to test whether or not the best-fit regions are significantly modified. 
No significant change is seen in the extent or shape of the highest K-S test probability regions in this case.

\subsection {Comparison}

The best-fit regions for the six pairs of parameters of the 
disk  population model have been displayed  and isoprobability coutours of the
K-S test drawn. These regions  are overlapping between the A- and FGK-type stars, 
indicating that there are best-fit values  
 that are compatible with both {\it Spitzer} samples. 
The fit values found in previous studies  for the A~stars \citep{Wyat07b} and 
for the FGK stars \citep{Kain11} differ significantly  
and may possibly be considered inconsistent. However, as is apparent in Figs~\ref{fig:Su06} and \ref{fig:Tril08}, 
they are specific solutions among an ensemble.  

\section {Steady state collisional evolution of debris disks around M dwarfs } \label {Mdwarfs}

\null

Prior to {\it Herschel}, two searches for cold debris disks around M~dwarfs  
were conducted in the far-infrared and at longer wavelengths. 
With {\it Spitzer}, \citet{Gaut07} observed 41 M~dwarfs at $70~\mu$m with a $1\sigma$ sensitivity 
level of $\sim 3$~mJy. Their sample is made of the nearest stars at less than 5 pc,  
likely older than 1Gyr, and of spectral types between M0 and M6.5. 
With the IRAM~30-meter and JCMT radiotelescopes, 
\citet{Lest06, Lest09} observed  50 M dwarfs at $\lambda=850~\mu$m and 1.2~mm with 
a $1\sigma$ sensitivity of 1-3 mJy. Half of this sample is made of  nearby stars at less than 7.7 pc
and likely older than 1~Gyr. The others are members of moving groups that have known ages of less than 
600~Myr, but are located farther away.

We have investigated whether or not the lack of detected debris disks around these M~dwarfs  
is consistent with the steady state collisional evolution model fit to the A and FGK data in the 
previous section.  This model is based on the assumption that 
the initial disk mass distribution is bimodal; only the high-mass collisionally-dominated 
disks are detectable, representing $\sim 25$~\% of the AFGK star sample,  
the others are low-mass PR-dominated disks, which are undetected and not used in the fit.

Thus, in our simulation of the  M~stars now, we used the predicted
flux density excesses from only 25 \% of the whole sample, just as we did for the AFGK stars. Hence,
for the \citet{Gaut07} survey, we picked  10-combinations out of the 35 stars 
with 70~$\mu$m measurements uncontaminated by cirrus or background galaxies, 
and  tested 100 such combinations to establish the statistics of the flux density excesses.

We compute the optically thin dust emission of a disk by integrating over the radial
distribution of the grains :

\begin{eqnarray}
  S_{\nu} & =  & {2.5~10^{11}} \times {{f_T^{-4}\times A(t_{age}) \times (\alpha+2)} \over {d^2 \times (r_{out}^{\alpha+2}-r_{in}^{\alpha+2})}} \times \nonumber \\
 & & {\sum_{i=1}^{N} B_{\nu}(T(r_i)) \times r_i^{\alpha+1} \Delta r},
\label{S_nu}
\end{eqnarray}

\noindent where  $S_{\nu}$ is in mJy, the distance $d$ to the star is in pc, the emitting surface of the dust
$A(t_{age})$ is in AU$^2$. We assume that small grains in these mostly old M~dwarfs are 
not blown out by stellar wind and adopted the minimum grain size of $D_{min}=0.1~\mu$m 
to estimate this surface, as discussed above. The  radial distance $r_i$,  
ranges  from the inner radius $r_{in}$ to the outer radius $r_{out}$, and is in AU, 
as the increment $\Delta r$. The  radial profile of the emitting cross-sectional area of the grains per unit area of the disk surface
is a power law with the standard index $\alpha$, as already described for the fractional 
dust luminosity in Eq~(\ref{eq:fd2}). The Planck function $B_{\nu}$ 
depends on the dust temperature  $T(r_i)$ at radial distance $r_i$. Several studies have shown that 
this temperature is higher than the blackbody temperature $T_{BB}(r_i)$  when  fitting simultaneously SED and images 
of resolved debris disks ; this is due to inefficient long wavelength emission
of small grains, which become hotter than blackbodies \citep[e.g.,][]{Back93,Lebr12}. 
A simplified model to account for these properties is $T(r_i) = f_T \times T_{BB}(r_i)$ where  $T_{BB}(r_i)=278 \times  L_*^{0.25} r_i^{-0.5}$ 
($L_*$ in L$_\odot$ and $r_i$ in AU). The factor $f_T$ 
is related to the grain emission efficiency $Q <1$ in a simple first-order physical interpretation. 
This factor has only been determined in a few cases~; 
$f_T=3.5^{+0.5}_{-1.0}$ for the M3~star GJ581 \citep{Lest12},
$f_T=1.9$ for the G star 61~Vir \citep{Wyat12}, and $f_T=1.4$  
for A stars  \citep{Bons10,Boot13}. This trend might be an indication that $f_T$ depends
on the stellar spectral type (higher $f_T$ for later type),  
although this is a tentative conjecture given the limited observations. 
Note that the effect of $f_T$  in Eq~(\ref{S_nu}) is double edges~; it
makes the Planck function peak toward shorter wavelengths, but at the same time the
factor $f_T^{-4}$ drasticly damps the flux density with $f_T > 1$. The overall effect 
on the flux density can go both ways depending on the Wien law maximum wavelength relative
to the observing wavelength. We tested both $f_T=1$ and  $f_T = 3.5$ in this study of M~dwarfs. 

From the best-fit region of the A star disk population in Fig~\ref{fig:Su06}, 
we adopt the smallest best-fit value for $D_c$, 2~km, the corresponding median mass, $M_{mid}=15~M_{\oplus}$,    
and the lower limits, $R1=2$~AU, and $R2=30$~AU. These parameters make the shortest collisional timescale 
$t_c(0)$  (Eq~\ref{eq:tc0}), and therefore the lowest flux densities at the age of the star $t$ 
(Eqs~\ref{eq:A},~\ref{S_nu}), as hinted by the observations in \citet{Gaut07}.
 
With these parameters, we simulate total masses $M_{tot}(t_{age})$ and mean disk radii $R$
 for the population of disks around the M dwarfs observed in the survey of \citet{Gaut07} and estimate their 
flux densities at $\lambda=70~\mu$m with Eq~(\ref{S_nu}). 
Their distances are known and their ages, which are undetermined but likely greater than 1~Gyr, were
chosen randomly between 1 and 10~Gyr. In selecting one hundred 10-combinations among the 35 M~dwarfs of the Gautier et al.'s sample,
and using the $3\sigma$ detection limit of their survey (9~mJy), 
the mean disk frequency is $5 \pm 3.5 $ \% ($f_T=1$) or $6 \pm 4 $\% ($f_T=3.5$) with these simulations.  
These two frequencies can be compared to the limit of 1\% set by their survey in which no disk is detected.
This limit is derived with the binomial discrete probability distribution 
$(Prob(X=k)=C_N^k p^k (1-p)^{N-k})$  used to quantify the chance of observing no disk ($k=0$) 
in a sample of $N=35$ stars when the mean detection 
rate is $p=1$\% and  the probability $Prob(X=0)$ is 68\% (equivalent to the $1~\sigma$ Gaussian 
limit). The higher disk frequency $p \sim 5\%$, found with our simulations above,
 corresponds to a probability $Prob(X=0)$ of 16\% to find no disk  in a sample of  $N=35$ stars.
This probability  is still less than the $2\sigma$ limit used in Gaussian statistics, 
and so the  disk evolution model fit to the AFGK stars 
can be considered consistent with the lack of disks detected around the M~dwarfs of the Gautier's sample.

We treated the submillimeter survey of M~dwarfs observed in \citet{Lest06, Lest09} 
the same way after correcting
the emission model by a factor $(210\mu$m/$\lambda)^{\beta}$ and using $\beta=1$ to account 
for the modified blackbody law at the longer observing wavelengths $\lambda=1200~\mu$m 
and $850~\mu$m. The $3\sigma$ sensitivity threshold is 3-9~mJy for this survey. 
The optimum model parameters ($M_{mid}=15~M_{\oplus}$, $D_c=2$~km, $R1=2$~AU, and  $R1=30$~AU) 
yield simulated detection frequencies of 0\%  for $f_T=3.5$ and only $\sim 2\% \pm 1.7$ for $f_T=1.0$. 
Note, surprisingly at first, that the frequency with  $f_T=1$ is higher than the frequency with $f_T=3.5$
because  the observing wavelength
$\lambda=1200~\mu$m is  longer than the Wien law maximum wavelength 
for the disk temperature : 
the increase of the Planck function when $f_T=3.5$ is counterbalanced by the factor  $f_T^{-4}$ 
in the flux density estimation. Nonetheless, the   
disk evolution model found for the AFGK stars is also consistent with the lack of disks 
detected around M~dwarfs at 850 and 1200 $\mu$m in the Lestrade et al. sample.

\section {Discussion}

\subsection { Comparison between the Kuiper belt and the debris disks around A and FGK stars} 

The best-fit values for the diameter $D_c$ of the  largest planetesimals range from 2 to 60~km
consistently for disks around both A- and FGK-type stars in Figs~\ref{fig:Su06} and \ref{fig:Tril08}, 
for our standard solution of broad disks (width $\Delta R=R/2$). This range extends from 6~km to
 100~km  for narrow disks ($\Delta R=R/10$) in  Fig~\ref{fig:Su06_bis}. 
It is remarkable that the size of the largest bodies of the collisional cascade in 
our modeled disk population  is close to  the break found at around 30~km in the Kuiper Belt 
Objects size distribution \citep{Fras09,Schl09,Fuen10}.
This break is usually attributed to the erosion of planetesimals caused by the
collisional evolution of bodies with size of $ < 30$~km over the age
of the solar system while larger objects
are primordial, {\it i.e.,} not yet colliding \citep{Schl13}. In the surveys used in our study, the mean age is 250 Myr for the  A stars
(about 1/3 of the lifetime for this
stellar type) and is 4.3 Gyr for the FGK stars, {\it i.e.,} close to the Sun age at about half of its 
lifetime. Although the Kuiper belt is presently a PR-dominated debris disk, 
it is thought that its mass was originally 
much higher than  its present estimate of $\sim0.1$~M$_{\oplus}$, possibly because of  
a catastrophic event in the meantime, such as the Late Heavy Bombardment 
about 700 Myr after the birth of the Sun  \citep{Gome05, Morb05, Tsig05}. 
Hence, the break at 30~km seen today is a scar of the past high collisional activity of 
the Kuiper Belt.  We have found a new similarity between
cold debris disks and the Kuiper Belt, which might be rooted in their common formation and evolution 
mechanisms. 
It has been shown that the large KBO size distribution can be well matched by
planet formation models of runaway growth where only a small fraction of the total mass
is converted into large protoplanets and most of the initial mass remains in 
planetesimals that are 1-10 km rather than $>100$~km in radius \citep[e.g.,][]{Keny99, Keny04, 
Schl11, Keny12}. 
A caveat is that we assume in our model that all disks  have the same $D_c$, 
and this is a limitation of our study. 

The best-fit value for the median initial mass $M_{mid}$ of debris disks around  solar type stars 
is $> 0.8$~M$_{\oplus}$ in our study and,
hence, is about ten times larger than for the present Kuiper Belt \citep{Glad01}. 
Consistently, \citet{Bryd06} conclude that cold debris disks detected around FGK stars have a median
fractional dust luminosity about ten times higher than the current solar value. 
These authors fit a theoretical Gaussian distribution 
in logarithmic space to the observed distribution of fractional dust luminosity frequencies 
by varying its median and standard deviation. 
    
\subsection{On the debris disks around M~dwarfs within the framework of our study}

We have shown that surveys have not been deep enough
to conclude that the disk population found with the steady state evolution model for  AFGK stars
is not consistent with the lack of disks detected around the M~dwarfs observed at far-infrared 
and submillimeter wavelengths.
A more stringent test is expected with the DEBRIS {\it Herschel} survey   because the sample is larger 
($N \sim 100$ for M-dwarfs) and the images are at least three times deeper than the past surveys.

\section{Conclusion}

In our study, we have fit the steady state collisional evolution model of  \citet{Wyat07a} to the
fractional dust luminosity distributions measured with {\it Spitzer} for the 
A and FGK stars. Hence, we have postulated that these distrubutions correspond to the population 
of high-mass collisionally-dominated disks that are detectable in the current surveys and
that can be analyzed {\it per se}. There is a distinct population of low-mass disks that are not considered
in the study because they are undetectable. The results of our study must be understood in this framework.

The model parametrized with $M_{mid}, D_c, R1$, and $R2$ satisfactorily fits  the observations
and the two resulting best-fit regions for the samples of  A and FGK stars 
are overlapping, indicating that there are solutions common to both samples. 

We have applied this steady state collisional evolution model to potential high-mass 
collisionally-dominated disks 
around M~dwarfs surveyed prior to the  {\it Herschel} satellite 
in the far-infrared by \citet{Gaut07} and in the submillimeter domain by \citet{Lest06, Lest09}.  
Within the framework of this study, we have shown that the disk population fit to the AFGK stars 
is still consistent 
with the lack of disks detected around M-dwarfs in the far-infrared and submillimeter surveys.

In a future study, we shall apply our novel approach  to study evolution of debris disks 
in the larger sample of M~dwarfs in the unbiased Herschel DEBRIS survey.

\begin{acknowledgements}
Etienne Morey's PhD work is funded by a grant from the Fondation CFM-JP Aguilar. We are in debt to our referee for his/her incisive comments which have greatly benefited 
to the paper.
This work is based in part on observations made with the Spitzer Space Telescope, which is operated by the Jet Propulsion Laboratory, California Institute of Technology under a contract with NASA.
\end{acknowledgements}

\bibliographystyle{aa}

%\bibliography{Etienne_13_v5}

%\begin{appendix} 

%\end{appendix}

\end{document}